# The simulation of multilayer magnetic films

V.Yu. Kapitan [1,2*], A.V. Perzhu[1], K.V. Nefedev[1,2]

[1] *Department of Computer Systems, School of Natural Sciences, Far Eastern Federal University, Vladivostok, 690950, 8 Sukhanova St., Russian Federation*
[2] *Institute of Applied Mathematics, Far Eastern Branch, Russian Academy of Science, Vladivostok, 690041, 7 Radio St., Russian Federation*

The results of numerical calculations of the thermodynamic properties of multilayer films with alternating magnetic and non-magnetic layers were presented. Computer simulation of such structures within the frame of the classical Heisenberg model was carried out by Monte Carlo methods. The processes of magnetization reversal of multilayer structures in external magnetic fields were investigated.

## 1. INTRODUCTION

The necessity of studying magnetic multilayer structures is explained by the prospects of their practical application as a technological base for creating new storages medium. Magnetic multilayer films have many unique features that contribute to increasing the density of information recording and the speed of storage devices. The energy-efficient data storages with ultra-high density recording is the next stage in the development of storage devices.

Multilayer structures are structures of alternating magnetic and nonmagnetic layers. In this connection, intensive theoretical and experimental studies of the properties of multilayer structures, depending on the thickness of the magnetic and nonmagnetic layers, their number and concentration are carried out [1-4].

In condensed matter physics is increase of the influence of computer simulation, which allows the development of new methods for the construction of complex nanostructures and which promotes the discovery of new phenomena and materials [5-13]. The study of collective behavior of magnetic systems will allow to predict and manage the properties of new materials, and will also make a significant contribution to the development of fundamental knowledge about the nano-world and ongoing processes.

## 2. A DESCRIPTION OF APPROACHES AND MODELS

### 2.1 The model of multilayer structures

Multilayer structures are structures of alternating magnetic and nonmagnetic layers, see fig. 1. Within the frame of our model of the multilayer structure, the magnetic layer of the multilayer one has a size of $N \times N \times L$, and represents, the system of spins of Heisenberg placed in the nodes of a simple cubic lattice, in which the spin has up to 6 nearest neighbors in its film, as well as interact via the long-range direct exchange with all the spins belonging to the neighboring magnetic films. Non-magnetic layers reduce the energy of the pair interactions between the spins of different films, depending on their thickness. In the developed software, this effect is implemented by reducing the value of the exchange integral between films $J_{12}$. Interaction was ferromagnetic inside the film and it was antiferromagnetic between films.

---

* kapitan.vyu@dvfu.ru

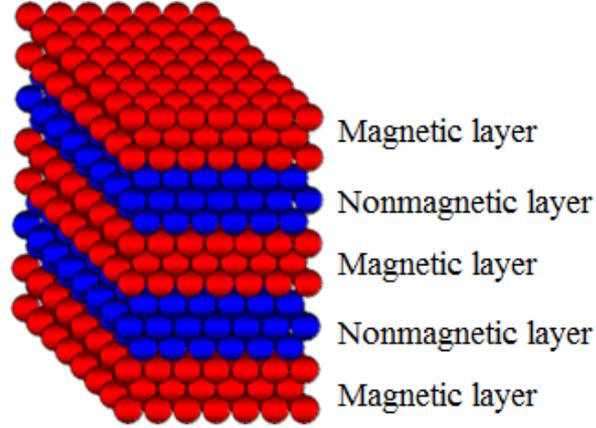

**Fig. 1** – The model of multilayer structures.

## 2.2  Heisenberg model

The Hamiltonian of the multilayers system in frame of Heisenberg model was set as follows:

$$H = -J_1 \sum_{\langle i,j \rangle} \vec{S}_i \vec{S}_j - J_{12} \sum_{\langle i,k \rangle} \vec{S}_i \vec{S}_k - A_z \sum_i \vec{S}_i - h_z \sum_i \vec{S}_i , \qquad (1)$$

where $\vec{S}_i$ is the atomic spin at the $i$-th lattice site, $J_1$ – ferromagnetic short-range exchange interaction inside each layer (it can be different for different layers), $J_{12}$ – antiferromagnetic long-range exchange interaction between neighboring layers, $|A|$ - constant of magnetic anisotropy, $h$ – external magnetic field. The sum is multiplied by 1/2 in order to avoid double summation, but some write it without the 1/2 factor; it's alright as long as one remembers that the summation is over $i$ and $j$ so that $i<j$.

A spin $\vec{S}_i = \{S_i^x, S_i^y, S_i^z\}$ is introduced as a three-dimensional unit vector in accordance with formulas (2 - 4) and see figure 2:

$$S_i^x = \sin(\varphi)\cos(\theta), \qquad (2)$$
$$S_i^y = \sin(\varphi)\sin(\theta), \qquad (3)$$
$$S_i^z = \cos(\varphi), \qquad (4)$$
$$\varphi \in [0,\pi], \theta \in [0,2\pi]$$

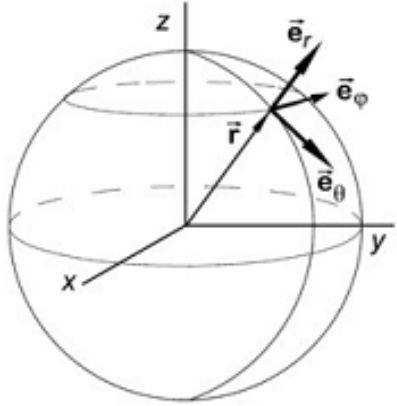

**Fig. 2** – A representation of a spin in a frame of Heisenberg model.

One may note that in the case of ferromagnets the above Heisenberg Hamiltonian predicts the parallel alignment of atomic spins, but doesn't specify a preferential direction of alignment, see the first two terms in Eq. 1. Thus, in such case the Heisenberg Hamiltonian is called the isotropic Heisenberg Hamiltonian. However, in real crystals the isotropy is broken by other magnetic effects that were neglected in the original Hamiltonian, like the dipolar interactions and spin-orbit coupling. Also, an external magnetic field can be applied so that the isotropy is broken by introducing a certain direction (the direction of the field). Thus, the whole Eq. 1 describes the anisotropic Heisenberg Hamiltonian.

The results obtained by the created software was verified in comparison with known results for the Heisenberg model [14], e.g. convergence of critical temperature $T_c = 1.44(3)$ has been reached.

### 2.3 Realization

The developed software is based on the new, promising programming language Rust. Rust is a new experimental programming language developed by Mozilla. The language is compiled, does not contain a garbage collector and is positioned as an alternative to C and C++. It is suitable for system programming and is comparable in speed and capabilities with C++, but it provides more security when working with memory. Rust supports functional, parallel, procedural and object-oriented programming, that is, almost the entire range of paradigms actually used in applied programming. From the above, it could be concluded that this language combines all the properties needed for numerical simulation. Also this language supports MPI for high performance computing.

For the Monte Carlo simulation, the Metropolis algorithm [15] and its parallel implementation using MPI and the Wang-Landau algorithm [16] were used. The calculation error for the Monte Carlo simulation, computed by the standard deviation, did not exceed 10%.

## 3. DISCUSSION.

### 3.1 Results

Within the frame of our model of the multilayer structure has from 3 to 5 of a magnetic layer and the magnetic layer of the multilayer one has a size of *10×10×3*, and represents, the system of spins of Heisenberg placed in the nodes of a simple cubic lattice.

The thermodynamic characteristics of multilayer structures, such as the temperature behavior of magnetization, energy, and heat capacity, were investigated using by the Monte Carlo methods, see Fig. 3.

In figure 3 (a), the maximum of magnetization did not reach 1 because in this case there were 3 layers in the multilayer film and the interlayer interaction was equal to $J_{1,2} = -0.3$.

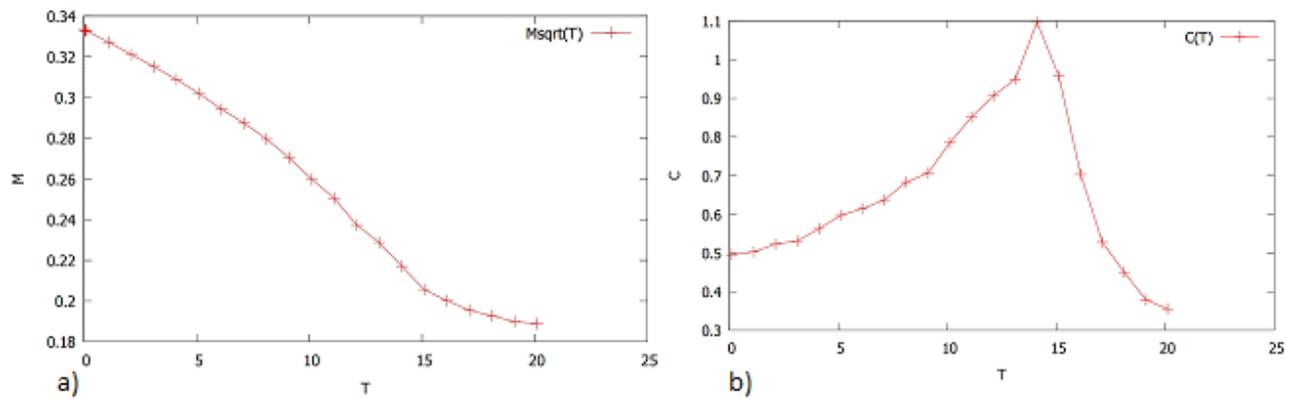

**Fig. 3** – The plots of temperature dependences for multilayer film: a) the quadratic magnetization, b) the energy

In figure 4 (a-d) was shown hysteresis loops for magnetic multilayer films with a different number of magnetic films. In fig. 4 (a) and (b) were shown behavior of the form of hysteresis loop depending on a number of MC steps for computer simulation, i.e. the time of the numerical experiment: in case (b) it was 10 times more than (a).

As could be seen from figures 4(b – d) for a different number of magnetic layers, the hysteresis loop behaves differently. If the film has an odd number of layers, an opening of the loop is observed at zero external magnetic field, due to the magnetization reversal of the middle layers earlier than the outside ones, because the presence of antiferromagnetic exchange between films.

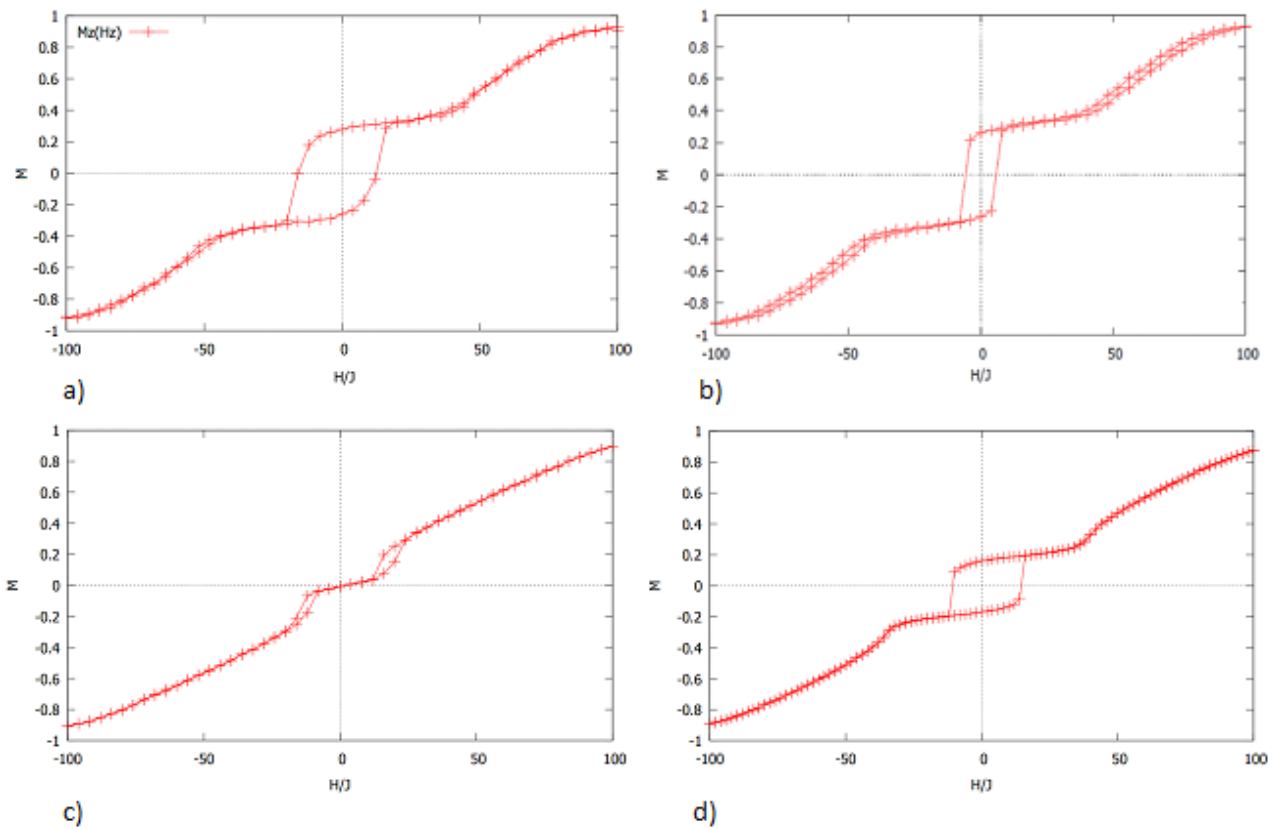

**Fig. 4** – The plots of hysteresis loop for multilayer films with a different number of magnetic films: (a, b) 3 magnetic films, c) 4 magnetic films, d) 5 magnetic films.

In the case of an even number of layers, see Fig. 4c, a collapsed loop is observed at zero magnetic field, due to the antiferromagnetic ordering of the films to each other, i.e. they compensate each other. The opening of the loops on the right and left sides are explained by the processes of their magnetization reversal over the field.

### 3.2 The further development of this study

The computer logic based on nanoscale magnetic structures will determine in the long-term future of computer technology, for the development of which will be used collective magnetic phenomena in the next generation information medium. Quite recently a new class of magnetic objects was discovered: magnetic skyrmions - vortex nanoscale long-lived magnetic formations. Originally, they were predicted theoretically [17, 18], and a little later they were experimentally discovered in bulk ferromagnetic materials with a crystal structure of B20 type [19]. The main magnetic state of such crystals is the spin helix, which arises from the presence of a special type of exchange interaction between the spins of neighboring atoms: the Dzyaloshinskii–Moriya (DM) interaction [20, 21]. In contrast to the usual exchange, which builds the neighboring spins collinearly, the DM exchange directs them perpendicular to each other. Subsequently, the skyrmions were detected in thin multilayer films, where the DM interaction is the result of symmetry breaking at the boundary of the magnetic and nonmagnetic layers [22]. Due to its topological nature, nanoscale and extremely low electric current required for the movement of skyrmions, they are proposed as elements for creating promising data storages will be operating on new principles - non-volatile memory and magnetic logic [23, 24]. Although a whole class of bulk crystals of a certain symmetry is found in which stable vortex formations are observed, by means of volumetric methods of investigation (neutron scattering) or local observation methods (transmission electron microscopy), the practical realization of memory based on magnetic skyrmions requires the use of technologies compatible with modern industrial production. In this regard, multilayer films are of particular interest. Recent impressive demonstrations of control of individual nanoscale skyrmions, including their creation, detection, manipulation and annihilation, have increased the probability of their use in future spintronics devices, including magnetic storage devices and magnetic logic [24].

### 4. CONCLUSION

In the frame of the classical Heisenberg model, lattice spin systems with direct short- and long-range exchange interactions were investigated by Monte Carlo methods. The researched systems are multilayer films with alternating magnetic and non-magnetic layers. The thermodynamic properties of such systems were investigated. Also, hysteresis phenomena were studied and the behavior of the hysteresis loop for various simulation parameters were considered. If the film has an odd number of layers, an opening of the loop is observed at zero external magnetic field, due to the magnetization reversal of the middle layers. In the case of an even number of layers, a collapsed loop is observed at zero magnetic field, due to the antiferromagnetic ordering of the films to each other, i.e. they compensate each other. The opening of the loops on the right and left sides are explained by the processes of their magnetization reversal over the field [25].

The developed software was based on the new, promising programming language Rust and MPI was used for parallelization.

### AKNOWLEDGEMENTS

Authors are grateful to Dr. Yurii Ivanov (Far Eastern Federal University, Vladivostok, Russia and Erich Schmid Institute of Materials Science Austrian Academy of Sciences, Austria) for his valuable comments.
This work was supported RFBR according to the research project No 16-32-00202, the Ministry